\documentclass[12pt]{article}

\usepackage{newtxtext,newtxmath}

\usepackage{graphicx}
\usepackage{xcolor}

\usepackage[letterpaper,margin=1in]{geometry}

\renewenvironment{abstract}
	{\quotation}
	{\endquotation}

\date{}

\makeatletter
\renewcommand{\fnum@figure}{\textbf{Figure \thefigure}}
\renewcommand{\fnum@table}{\textbf{Table \thetable}}
\makeatother

\usepackage{scicite}

\usepackage{url}

\def\scititle{
	Entropy of state transitions in macroscopic active matter
}
\title{\bfseries \boldmath \scititle}

\author{
	Francesco Roman\`o$^{1\ast\dagger}$ and
	Michael Riedl$^{2\ast\dagger}$\and
	\small$^{1}$Univ. Lille, CNRS, ONERA, Arts et Metiers Institute of Technology, Centrale Lille, UMR 9014 - \vspace{-0.25cm}\\ \small LMFL - Laboratoire de Mécanique des Fluides de Lille - Kampé de Fériet, F-59000 Lille, France.\and
	\small$^{2}$Physics of Life, TU Dresden, Dresden, 01307, Germany.\and
	\small$^\ast$Corresponding author. Email: francesco.romano@ensam.eu, michael.riedl@tu-dresden.de\and
	\small$^\dagger$These authors contributed equally to this work.
}

\begin{document} 

\maketitle

\begin{abstract} \bfseries \boldmath
The extension of thermodynamic principles to active matter remains a challenge due to the non-equilibrium nature inherent to active systems. 
In this study, we introduce a framework to assess entropy in our minimal macroscopic experiment based on the utilized degrees of freedom. 
Using motorized spheres as active particles, we demonstrate that the system transitions between distinct active states. Analogous to the phase transition in classic solids, liquids, and gases, each phase is characterized by a quantifiable change in entropy. 
We show that the corresponding phase transitions are accompanied by discrete jumps in entropy, resulting from newly utilized degrees of freedom.
Our findings reveal that active matter can exhibit phase transitions analogous to classical thermodynamic systems, quantifiable in terms of their entropy and temperature.
By bridging equilibrium thermodynamics and active matter, this work shows how underlying principles extend to complex, living systems.
\end{abstract}

\noindent

Active matter systems operate far from equilibrium and encompass living as well as inanimate entities.
Populations of these individuals show the remarkable ability to transition from autonomous movement to unified motion \cite{vicsekNovelTypePhase1995, devereuxEnvironmentalPathEntropyCollective2023}. 
Collectives exhibit these ordered states at vastly different scales. 
Examples can be found in bacterial colonies \cite{bricardEmergentVorticesPopulations2015, japaridzeSynchronizationColiBacteria2024, aubretTargetedAssemblySynchronization2018, groberUnconventionalColloidalAggregation2023}, cell populations \cite{doxzenGuidanceCollectiveCell2013, jainRoleSinglecellMechanical2020, riedlSynchronizationCollectivelyMoving2023}, flocks of birds \cite{cavagnaScalefreeCorrelationsStarling2010, cavagnaMarginalSpeedConfinement2022, portugalUpwashExploitationDownwash2014}, synthetic self-propelled particles \cite{kokotSpontaneousSelfpropulsionNonequilibrium2022, kokotManipulationEmergentVortices2018} and swarms of robots \cite{riedlSynchronizationCollectivelyMoving2023, chvykovLowRattlingPredictive2021,rubensteinProgrammableSelfassemblyThousandrobot2014a}. 
Intrinsic to each individual is an energy source that fuels the mechanism driving self-propulsion.
Over time, the consumed energy dissipates in the form of heat and cannot be recycled to generate further movement.
This continuous transfer of energy sets active systems apart from classical systems, where no net energy flux occurs and global equilibrium is maintained.
In classical systems a thermal bath acts as a reservoir, storing and supplying the energy that facilitates the motion of its constituents.
Although related, the differences between both fields render bridging concepts from thermodynamics to active systems a complex task 
\cite{hechtHowDefineTemperature2024,fodorHowFarEquilibrium2016,mandalEntropyProductionFluctuation2017,obyrneTimeIrreversibilityActive2022,dabelowThermodynamicNatureIrreversibility2023}.
Particularly in experimental active systems, assessing fundamental quantities like temperature and entropy is not straightforward \cite{tanOddDynamicsLiving2022,chvykovLowRattlingPredictive2021}.  

Entropy is a statistical measure based on the probability of observing a given microstate. 
In thermodynamics, a microstate refers to a specific configuration of all particles in a system, characterized by their positions and momenta at a particular instant.
Evaluating the probability of a certain microstate hinges on the conceptualization of every possible microstate, which cumulatively define the macrostate.
However, in active matter systems, the entirety of all potential microstates can typically neither be directly observed  nor defined a priori. 
Therefore, entropy can only be approximated based on the states observed, which renders its calculation in experimental systems challenging \cite{chvykovLowRattlingPredictive2021}. 
This limitation leaves a fundamental question unanswered: How far can we extend classical thermodynamic concepts to active matter?

\begin{figure}[t] 
	\centering
	\includegraphics[width=.8\textwidth]{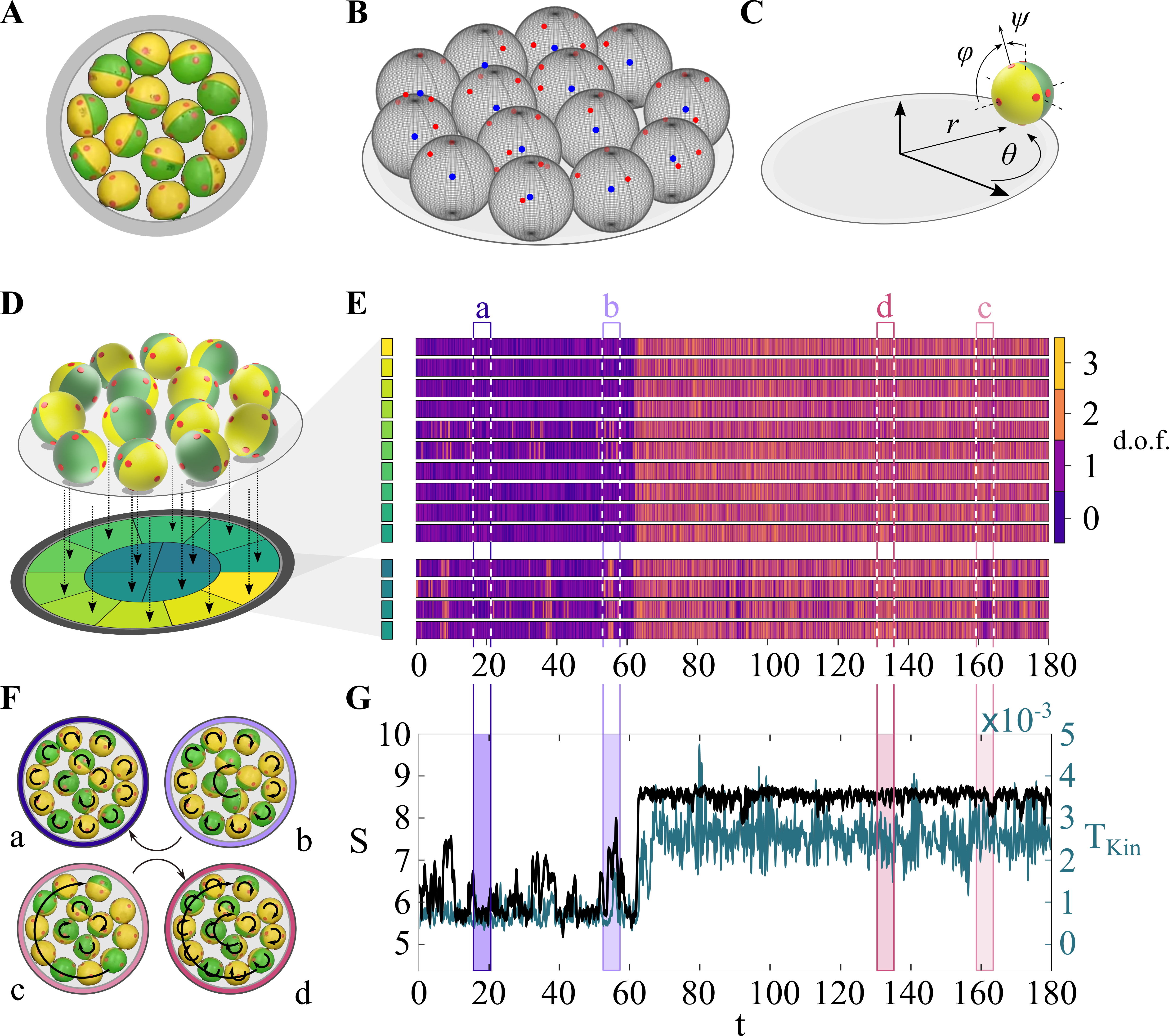} 

	\caption{\textbf{Self-organizing dynamics for the experimental active spheres system.}
		(\textbf{A}) Top-down view of motorized spheres confined within a cylindrical boundary. Motors rotate around an axis normal to the yellow-green hemisphere division. Red markers aid in post-processing. (\textbf{B}) Reconstruction of each sphere's instantaneous state, with spheres as wireframes showing the red markers (Panel A) and blue dots as geometric centers. (\textbf{C}) Three degrees of freedom: (i) radial motion $r$, (ii) azimuthal motion $\theta$, and (iii) localized spinning around the plane-normal sphere's axis. (\textbf{D}) Discretization of the domain into $N$ equal-area bins, with 14 sectors dividing the confinement. Arrows indicate projection of the properties of each sphere to the nearest bin. (\textbf{E}) Heatmap showing degrees of freedom over time. Each bin is in one of five states: (i) empty, (ii) zero, (iii) one, (iv) two, or (v) three degrees of freedom. In this configuration, every bin corresponds to a sphere with specific degrees of freedom, no bin is in the `empty' state. (\textbf{F}) Four system states in a fully packed configuration: (a) only spinning is active; (b) outer ring motion is uncorrelated, inner ring synchronizes azimuthally; (c) the inverse of (b); (d) both rings are synchronized with two degree of freedom per sphere. (\textbf{G}) Time series of entropy $S$ (black) and kinetic temperature $T_{\text{kin}}$ (teal). Vertical bands indicate key transitions corresponding to Panel E, identifying the system states depicted in Panel F. 
 }
	\label{fig:figure_1} 
\end{figure}

Increasing entropy in a classical system provides a clear direction for the progression of time.
This is not the case for active systems and the arrow of time becomes ambiguous.
We encountered this ambiguity while studying phase transitions in a minimal system of actives spheres \cite{riedlSynchronizationCollectivelyMoving2023}.
These autonomously moving spherical shells are driven by an internal, unbalanced motor \cite{riedlSynchronizationCollectivelyMoving2023,prentisExperimentsStatisticalMechanics2000}. 
The motor applies a constant torque on an internal shaft along an axis connecting two poles of the sphere. 
While the shell is fixed, the applied torque causes the continuous revolution of the internal mass. 
When releasing the sphere on a sufficiently rough substrate, this mechanism causes the sphere to roll along a chaotic trajectory. 
A population of these motorized spheres can transition from a disordered motion to a self-organized, collectively rotating pattern \cite{okeeffeOscillatorsThatSync2017,riedlSynchronizationCollectivelyMoving2023}; a transition driven purely by mechanical interactions  (fig.\ \ref{fig:figure_1}A, movie 1\ \ref{movie:S1}A).
We observed an intriguing feature of this system.
The system shows an intriguing feature: In a stable state, the dynamics are time-reversible, making it impossible to distinguish, at first glance, whether a recording is played forward or backward (movie 1\ \ref{movie:S1}B, movie 1\ \ref{movie:S1}C).
The system’s dynamics appear the same in either temporal direction; a feature that is lost during a phase transition (movie 1\ \ref{movie:S1}D). 
Since the progression of time in our system appears unclear, we aim to quantify this loss of temporal symmetry in terms of  entropy.

We define entropy based on the degrees of freedom utilized within our system. 
To extract these degrees of freedom, we label each spherical shell (diameter $d = 84$\,mm) with several red markers (fig.\ \ref{fig:figure_1}A). 
Over the course of our experiments, we track the location  of each shell and its corresponding markers independently within a circular confinement at a frame rate of $30$\,Hz (movie 2\ \ref{movie:S2}A).
The confinement diameter $D$ spans $380$\,mm, approximating to slightly more than $4d$. We chose $D$ to accommodate a packed configuration of $N_s = 14$ spheres, while allowing enough space to permit localized spinning of each individual. 
Based on these data, we reconstruct a three-dimensional instantaneous representation of our experiments (fig.\ \ref{fig:figure_1}B, movie 2\ \ref{movie:S2}A). 
This reconstruction allows us to decompose the motion of each shell into its translational and rotational components.
A sphere typically has six degrees of freedom: three translational and three rotational. 
Our active spheres are limited to a two-dimensional translation, which reduces the translational degrees of freedom to two. 
These are defined with respect to the center of the arena: one describing radial motion (toward or away from the center) and the other describing azimuthal motion (around the center) (fig.\ \ref{fig:figure_1}C).
The rotational degrees of freedom are characterized by the displacement of the red markers fixed to each spherical shell. 
Since the markers are constrained to the radius of the shell, one rotational degree of freedom is lost.
This leaves two rotational degrees of freedom, defined by two angular coordinates in a spherical coordinate system (fig.\ \ref{fig:figure_1}C).  
We calculate the magnitude of the two angular velocities to quantify the overall rotational speed. 
This value in speed indicates whether the rotational degree of freedom is being active, omitting the direction of rotation. 
We define a threshold value that allows to distinguish whether the corresponding degree of freedom is active and validate the robustness of our results to variations its value.
This consideration leaves us with a total of three degrees of freedom that each sphere can utilize, the two translational degrees of freedom and one spinning degree of freedom. 

Since these degrees of freedom are fully accessible in our minimal experimental system and can be used to define each state, we can now focus on how to calculate entropy from them.
To calculate the entropy level $S$ of a state in an isolated system, one needs to compute all potential configurations $\Omega$ while keeping the energy of the system constant. 
To generalize this concept to an active system, we need to preserve the degrees of freedom associated to each sphere.
We therefore assume that the potential configurations are all equally probable. 
This represents an assumption analogue to the kinetic theory underlying thermodynamics, which also hypothesizes a uniform probability of occuring states. 
To compute all potential configurations, we discretize the confinement area into $N$ bins of equal area (fig.\ \ref{fig:figure_1}D). 
The number of bins corresponds to the highest number of spheres that can be theoretically fitted in the available space, approximating the configuration with the highest possible packing ratio. 
For each of these bins, five possible states can be assigned. 
The bin is either: (i) empty, or filled with an active sphere having (ii) zero, (iii) one, (iv) two, or (v) three degrees of freedom. 
In the presented experiments, we fix the confinement and sphere sizes while changing the number of spheres. 
Therefore, a bin can only be in an `empty state' (i) if less than 14 individuals are employed.
We project the utilized degrees of freedom of each sphere onto the bin it occupies. 
Representing the bins in terms of their radial distance from the center distinguishes an inner and outer ring. 
Collectively, the degrees of freedom represented this way represent each state transition over time in our experimental system (fig.\ \ref{fig:figure_1}E, movie 2\ \ref{movie:S2}B). 

Representing the experimental dynamics in terms of utilized degrees of freedom allows us to identify four distinct states in the fully packed configuration (fig.\ \ref{fig:figure_1}F, movie 2\ \ref{movie:S2}B): state \textbf{a}, characterized by frequent collisions with little net displacement, using solely the localized spinning degree of freedom; state \textbf{b}, the motion of the spheres on the outer ring remains uncorrelated, the inner ring synchronizes azimuthally and achieves an additional degree of freedom per sphere; state \textbf{c}, inverted to the previous state, showing an uncorrelated inner ring and a synchronized outer ring; and state \textbf{d}, where both rings are synchronized. While states \textbf{a} and \textbf{d} represent stable configurations that persist over long periods, states \textbf{b} and \textbf{c} are unstable. The unstable configurations \textbf{b} and \textbf{c} appear only transiently, while triggering a global transition towards \textbf{a} and \textbf{d}, respectively. 

The utilized degrees of freedom projected on the discretized experimental area lets us compute all possible configurations $\Omega$. 
The general formulation of $\Omega$ would assume that all the degrees of freedom can be allowed in each bin of the domain. 
In a fully-packed system, an active sphere initially occupying the outer ring will be forced to keep moving on the outer ring; its radial degree of freedom is unavailable. 
This volume excluding effect limits the number of configurations admissible. 
To take this into account, we evaluate $\Omega$ for each ring of bins independently ($\Omega_k$ for $k=1$, 2 in our two-ring-bin setup). 
Multiplying the possible configurations of the inner annulus $\Omega_1$ with the ones of the outer annulus $\Omega_2$, we compute all possible permutations $\Omega$:
\begin{equation}\label{eq:ConfigurationsGeneral}
    \Omega_k = \begin{pmatrix}
        e_k + \sum_{i=0}^I b_{i,k} \\ e_k
    \end{pmatrix} \Pi_{i=0}^I \begin{pmatrix}
        \sum_{j=i}^I b_{j,k} \\ b_{i,k}
    \end{pmatrix}(i+1)^{b_{i,k}}, \qquad \Omega = \prod_{k=1}^K \Omega_k,
\end{equation}
where $e_k$ is the number of empty bins in the $k-$th ring, $i$ denotes the number of degrees of freedom, $b_{i,k}$ the number of bins with $i$ degrees of freedom in the $k-$th ring, $I$ is the maximum amount of degrees of freedom ($I=3$ in our system), and $K$ the maximum amount of rings ($K = 2$ in our system). From this definition follows that the number of total bins $N$ is the sum of the empty bins $e_k$ and the occupied bins over the two rings, i.e. $N=\sum_{k=1}^K \left(e_k + \sum_{i=0}^I b_{i,k}\right)$. This approach leads to the entropy of the system $S = \log(\Omega) = \log(\Pi_{k=1}^K\Omega_k) = \sum_{k=1}^K\log(\Omega_k)$.
Evaluating entropy in our system based on degrees of freedom, shows that steady states are associated with a statistically constant value while a jump in entropy corresponds to the previously defined state transitions (fig.\ \ref{fig:figure_1}G).

In addition to the entropy we quantify the energy levels corresponding to the previously found states in terms of the kinetic temperature $T_\text{kin}$ \cite{hechtHowDefineTemperature2024, hechtMotilityinducedCoexistenceHot2024}. 
The instantaneous kinetic temperature is defined as the variance of the kinetic energy $T_{kin} = \frac{1}{2 \delta N_s} \sum_{n=1}^{N_s} \left(v_n - \langle v \rangle \right)^2$,
where $\langle v \rangle$ denotes the average over the population of $N_s$ spheres in their two dimensional space $\delta$ and $v_n$ is the translational velocity magnitude of the $n-$th sphere.
Using this definition, we can also characterize state transitions observed in our system as a function of the kinetic temperature (fig.\ \ref{fig:figure_1}G).
The average kinetic temperature levels are remarkably distinct with respect to temperature fluctuations, even for a limited amount of active spheres.
In summary, we demonstrate that a distinct separation between system states is experimentally found in kinetic temperature as well as  entropy (see fig.\ \ref{fig:figure_1}G, , movie 2\ \ref{movie:S2}B). 

\begin{figure} 
	\centering
	\includegraphics[width=1\textwidth]{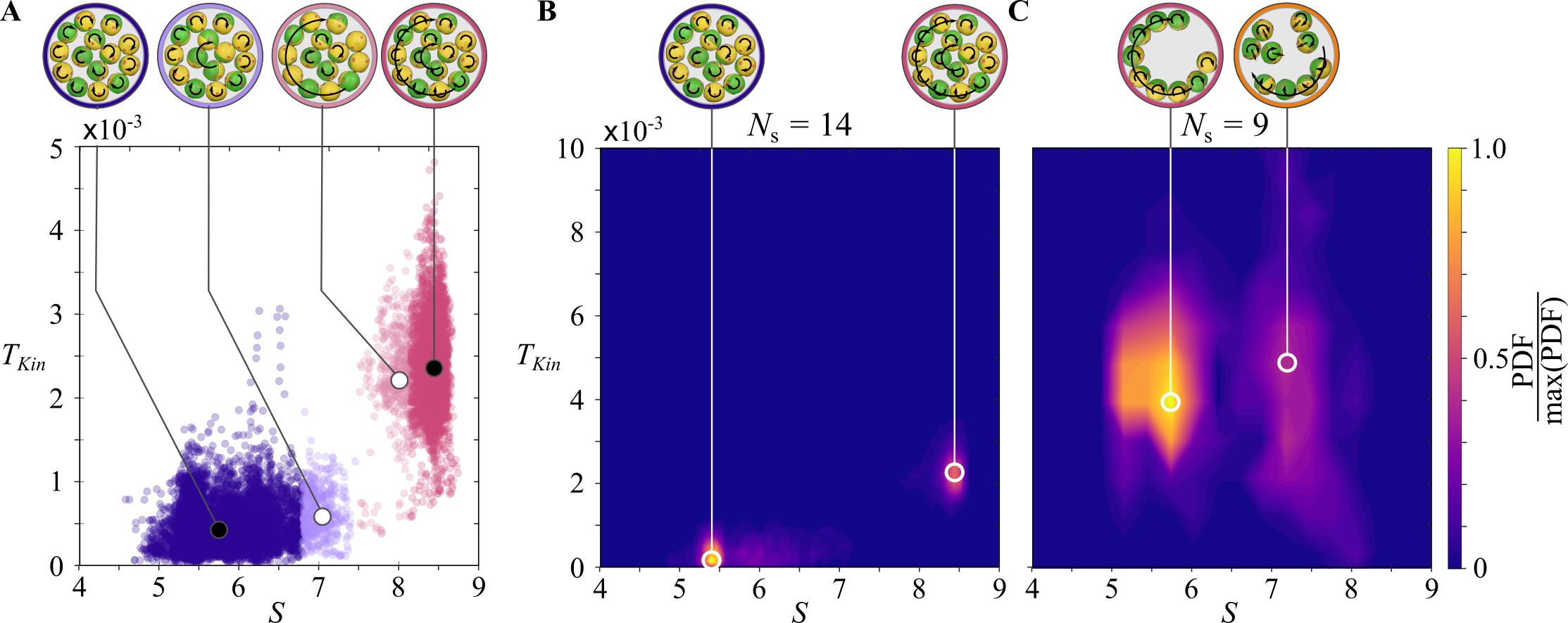} 

	\caption{\textbf{ State characterization and probability density analysis of the system in entropy $S$ and kinetic temperature $T_\text{kin}$.}
		(\textbf{A}) Each dot represents an instantaneous characterization of the system with 14 spheres (highest packing ratio) across multiple experiments, plotted in the entropy-kinetic temperature plane. Black markers indicate the centroid of stable states, corresponding to high values in the probability density function (PDF) plot of Panel B, while white markers denote transient states. The color coding of the markers and outer rings of the system states (see top row) corresponds to the temporal bands highlighted in Figure 1G, as well as the ring colors in Figure 1F. (\textbf{B}) Normalized PDF for the 14-sphere system revealing the likelihood of the system occupying specific states (i.e. state (a) and (d) of Figure 1F). The highest normalized PDF is associated to the two stable states depicted above. (\textbf{C}) Equivalent representation to Panel B, but for a system with 9 motorized spheres.}
	\label{fig:phase_space} 
\end{figure}

Representing the behavior of the system in terms of kinetic temperature and entropy captures the previously identified states, e.g. for $N_s=14$ (fig.\ \ref{fig:phase_space}A). 
The four states (a-d) are color-coded and depicted in fig.\ \ref{fig:phase_space}A and demonstrate the separation in entropy typical of thermodynamic systems upon the occurrence of a phase transition (see gap between (a,b) and (c,d)). 
The transition from unstable (b,c) to stable (a,d) states is indicated by an arrow. 
The stability of states (a) and (d) is demonstrated in fig.\ \ref{fig:phase_space}B by computing the probability density function (PDF) over time and over several repetition of the same experiment. 
The PDF shows that the system localizes within two foci centred in state (a) and (d), which points towards a bi-stable configuration. 
A similar behavior can be found upon a variation of $N_s$ for all our experiments. 
Normalizing the PDF by its highest value, $\max(PDF)$, we can visually compare it to other cases. 
For example in the case of $N_s=9$ in fig.\ \ref{fig:phase_space}C (movie 2\ \ref{movie:S2}C), the two stable states correspond to: ($\alpha$) a low-entropic, low-temperature state that synchronizes all its individuals along the outer ring, and ($\beta$) an high-entropic, high-temperature state that allows for chaotic motion of the spheres in the void region.  
This representation allows to visually distinguish individual states from each other and also indicates their stability.

Extending this approach to varying numbers of spheres $N_s$ in the system (movie 2\ \ref{movie:S2}C), we can quantify all potential states and transitions in terms of kinetic temperature and entropy. 
To compare systems with various number of spheres, we need to consider effects resulting from the statistically small number of individuals present. Therefore, we normalize the computed entropy by a reference value that accounts only for the permutations of the occupied bins. This reference value corresponds to the most stable configuration for the considered number of spheres. This most stable configuration represents an outer ring where each occupied bin has $i = 2$ degrees of freedom. 
This configuration persists until the outer ring is fully occupied, then the inner ring fills up, leading to $S_\text{ref} = \log(3^{\sum_i^I b_{i,1}}\times 3^{\sum_i^I b_{i,2}})$.
Normalizing each state entropy with respect to the corresponding reference entropy allows to compare configuration different number of active spheres.

\begin{figure} 
	\centering
	\includegraphics[width=0.8\textwidth]{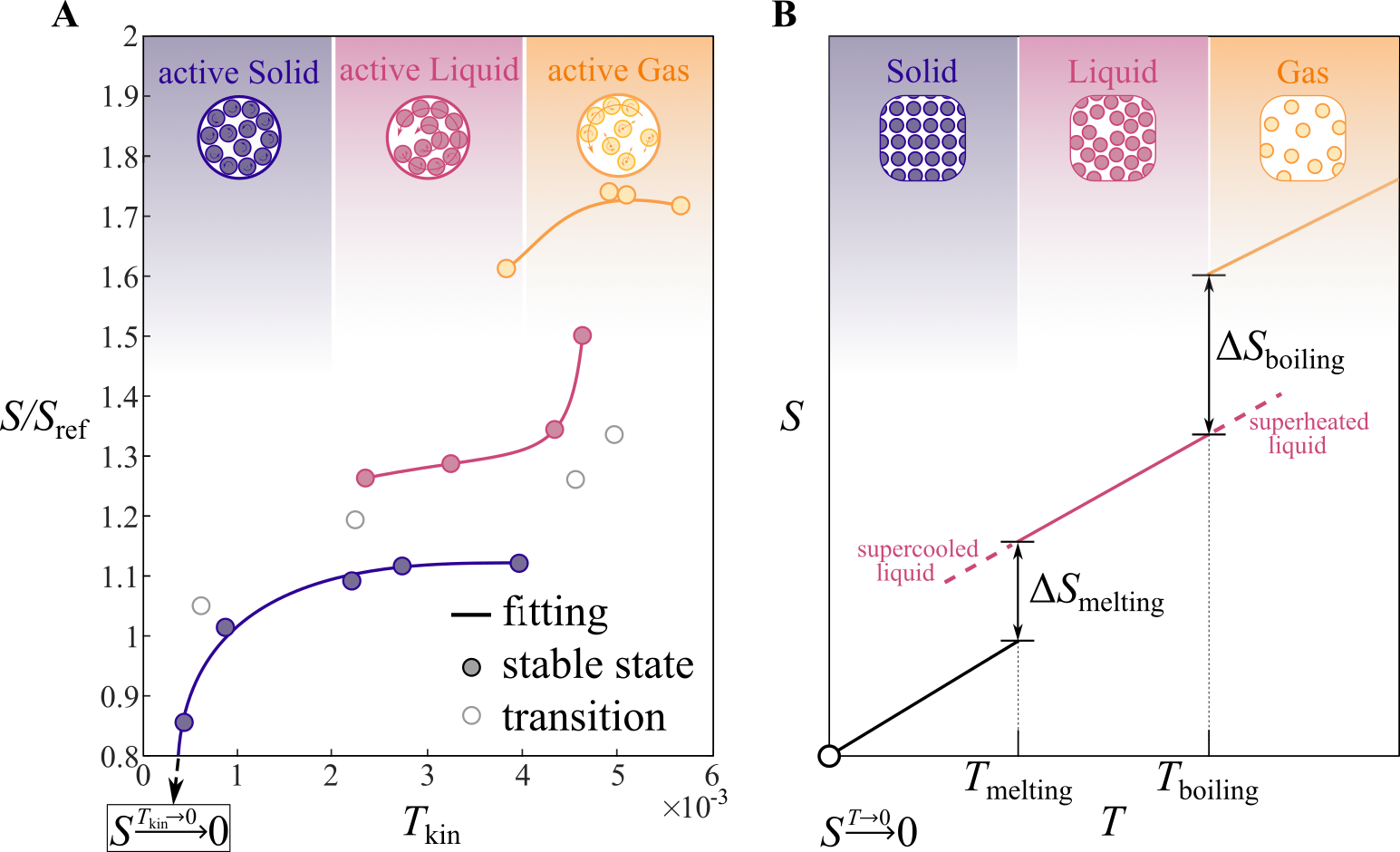} 

	\caption{\textbf{Analogy between state transitions in active systems and phase transitions in thermodynamic systems.} (\textbf{A}) Relationship between the system's normalized entropy $S/S_{\text{ref}}$ and kinetic temperature $T_{\text{kin}}$ across active solid, liquid, and gas states. The stable states (full circles) correspond to the black markers of Figure 2A obtained across several experiments with different amount of motorized spheres. The empty markers denote transient unstable states of the active system of motorized spheres (see white markers in Figure 2A). The fitted lines depict how the system evolves through different phases, with each region corresponding to a distinct state of matter, color-coded as in Figures 1F, 1G, and 2A. (\textbf{B}) The corresponding thermodynamic analogy highlights phase transitions, showing entropy $S$ as a function of temperature $T$ for solid, liquid, and gas states. Key thermodynamic phenomena such as supercooling, superheating, and the entropy changes associated to phase transition ($\Delta S_{\text{melting}}$ and $\Delta S_{\text{boiling}}$) are indicated. The color coding in Panel A aligns with the different states in Panel B, emphasizing the parallel between state transitions in active systems and classical phase transitions.
}
	\label{fig:phase-transition} 
\end{figure}

Gathering all the experiments in single representation, we depict the mean value of the kinetic temperature and the normalized entropy for each state identified in our experiments (fig.\ \ref{fig:phase-transition}A). We observe that three stable states are robust features of our system, regardless of the number of individuals. These states correspond to: an active solid, where spheres perform a localized spin while remaining uncorrelated; an active liquid, where spheres synchronize into azimuthally-propagating coherent structures, an active gas; where spheres move chaotically within the void area of the domain. Each of these regimes is separated by a sharp increase in entropy, analogous to what can be observed in thermodynamical systems for phase transitions between solid, liquid, and gas (fig.\ \ref{fig:phase-transition}B).
Thus, the representation of kinetic temperature and entropy not only captures these stable states but also highlights the fundamental parallels with classical phase transitions in thermodynamic systems.

In this analogy, the active solid state corresponds to a classic solid, where molecules primarily vibrate without significant net displacement. This leaves the thermodynamic solid state with the vibration of its constituents as its one predominant degree of freedom. The actives spheres in our system analogously achieve only the spinning degree of freedom, in the highly packed, uncorrelated active solid state \cite{aubretTargetedAssemblySynchronization2018, capriniEntroponsCollectiveExcitations2023,baconnierSelectiveCollectiveActuation2022,tanOddDynamicsLiving2022,golestanianBoseEinsteincondensationScalarActive2019,mengMagneticMicroswimmersExhibit2021}.  A solid loses this last degree of freedom of vibration only when two conditions are met: (i) a perfect crystalline structure (ii) at absolute-zero temperature. In our system, the non-synchronized state admits this zero-entropy state as well: if (i) the maximum packing ratio is attained and (ii) all the motors are switched off. 
This parallel between the active solid and classical solids underscores how limiting the degrees of freedom results in a distinct low-entropy state, demonstrating that even in non-equilibrium systems, structural constraints drive predictable thermodynamic behaviors.

Further continuing with the analogy, the active liquid state corresponds to a classical liquid phase. A thermodynamic liquid, beside vibrating like in a solid, admits molecules to translate more freely.
This increases their degrees of freedom with the dimensions of space. A positive jump in entropy occurs, when passing from a solid to a liquid phase, corresponding to availability of the newly accessible states. For our motorized sphere system, we identify the analogous behaviour, when the transition to coordinated movement occurs. The transition from uncorrelated active solid state to the synchronized active liquid state coincides with a jump in entropy and an increased kinetic energy of the system \cite{levisActiveBrownianEquation2017,paoluzziNoiseInducedPhaseSeparation2024,digregorioFullPhaseDiagram2018,hechtMotilityinducedCoexistenceHot2024}. In thermodynamics, melting represents an intermediate regime, where two phases are present; liquid and solid. We identified such a transition state of coexistence in our system as depicted in fig.\ \ref{fig:phase-transition}A. 
Extending our analogy to classical liquids, the transition from uncorrelated motion to synchronized movement in the active liquid phase is captured by the increasing entropy during phase transitions.

Finally, when considering the active gas state, we draw the analogy with classical gases, whose molecules are free to move; unlocking an even larger range of potential states (fig.\ \ref{fig:phase-transition}B). With a decreasing number of spheres this active gas phase arises in our system (fig.\ \ref{fig:phase-transition}A). Radial displacements occur with a higher probability and a chaotic signature emerges \cite{shankarHiddenEntropyProduction2018,hechtMotilityinducedCoexistenceHot2024,devereuxEnvironmentalPathEntropyCollective2023}. This behavior enables a larger number of potential states reflected in the corresponding jump in entropy, analogous to boiling where a liquid transitions to a gas. The additional degrees of freedom for the molecules increase, which promotes another positive jump in entropy. 
Like its classical counterpart, the active gas state demonstrates how increasing degrees of freedom lead to higher entropy.

We demonstrate that entropy can be defined and quantified in terms of the utilized degrees of freedom within a system. 
Using our minimal yet fully measurable experimental setup, we establish that fundamental thermodynamic principles can be applied to active macroscopic systems without losing their generality. 
Analogous to classical solids, liquids, and gases, our system of active spheres exhibits distinct states and phase transitions tied to changing degrees of freedom.
The minimal nature of our system allows us to map these dynamics through both entropy and kinetic temperature.
This reveals that the three active phases are characterized by distinct levels in entropy, including the emergence of transient states during phase transitions. 
These findings show that phase transitions in active systems at the macroscale exhibit analogue behavior to those observed in thermodynamic systems at the microscale.
Our results point towards a deeper, universal connection between thermodynamics and active matter, and our minimal system provides an ideal setup to explore these connections at a fundamental level. 
While it remains to be seen whether the presented analogy extends to living systems \cite{tanOddDynamicsLiving2022,riedlSynchronizationCollectivelyMoving2023}, the extension of thermodynamic concepts to active systems through minimal and controllable experimental setups offers a promising path for advancing our understanding of state transitions across scales in active matter.

\newpage
\clearpage

\clearpage 

%
\bibliography{science_template} 
\bibliographystyle{sciencemag}

%
%
%
%
%
%


\section*{Acknowledgments}
We thank Jan Brugués, Matthew Turner, Claire Dessalles, Nico Schramma and Astrid Riedl-Fajtak for their valuable feedback, insightful discussions, and help with proofreading, which greatly contributed to the development of this work.

\paragraph*{Funding:}
This research received no external funding.

\paragraph*{Author contributions:}
Both authors contributed equally to design of the study, conducting the experiments, developing the theoretic aspects of this work and writing of the paper. 

\paragraph*{Competing interests:}
There are no competing interests to declare.

\paragraph*{Data and materials availability:}
The authors declare that the data supporting the findings of this study are available within the paper and its supplementary information files. Imaging data sets generated during the current study are available from the corresponding authors on request.



\newpage


\renewcommand{\thefigure}{S\arabic{figure}}
\renewcommand{\thetable}{S\arabic{table}}
\renewcommand{\theequation}{S\arabic{equation}}
\renewcommand{\thepage}{S\arabic{page}}
\setcounter{figure}{0}
\setcounter{table}{0}
\setcounter{equation}{0}
\setcounter{page}{1} 


\begin{center}
\section*{Supplementary Materials for\\ \scititle}

Francesco Roman\`o$^{1\ast\dagger}$ and
Michael Riedl$^{2\ast\dagger}$ \\
\small$^\ast$Corresponding authors. 
\\Email: francesco.romano@ensam.eu, michael.riedl@tu-dresden.de\\
\small$^\dagger$These authors contributed equally to this work.
\end{center}

\subsubsection*{This PDF file includes:}
Materials and Methods\\
Captions for Movies S1 to S2\\

\subsubsection*{Other Supplementary Materials for this manuscript:}
Movies S1 to S2\\

\newpage


\subsection*{Materials and Methods}

\subsubsection*{Active sphere preparation}
Upon receiving the motorized spheres (D.Y. Toy), we removed them from their packaging and detached the weasels attached to the shell. Each sphere consists of two screwable half-spheres, which can be opened to insert a single AA battery. Before each experiment, we replaced all AA batteries to ensure consistent torque generation across the entire population. Failing to do so, especially with low or depleted batteries, could introduce artifacts in the rolling behavior of the spheres. The confinements used in the experiments were 3D printed using PLA filaments.

\subsubsection*{Image analysis}
Fiji imaging processing software was used for the processing and analysis of images and video microscopy data \cite{schindelinFijiOpensourcePlatform2012a}.

\subsubsection*{Tracking and postprocessing}
To extract the trajectories of the spheres and dots from recorded videos, we used the TrackMate plugin within Fiji \cite{tinevezTrackMateOpenExtensible2017a}. The spheres and dots can be separated by color thresholding, leading to two sets of images. Each set is tracked individually and post-processed using a script in Python 3.7 or Matlab. Both sets of tracks are sufficient to create a reconstruction of the system.

\clearpage 

\textbf{Time symmetry during states and transitions in a system of motorized spheres}
\textbf{A} Self-organization of motorized spheres. This video shows the emergence of collective rotational motion in a population of motorized spheres confined in a circular domain. Initially, the spheres move chaotically without correlation. Over time, the spheres begin to synchronize their movement, forming coherent structures akin to an active liquid phase.
\textbf{B-C} The progression of time in both the chaotic and ordered states is time-reversible. Distinguishing whether a  movie is being played forward or backward becomes difficult.
\textbf{D} The progression of time during state transitions. During state transitions time reversibility symmetry is broken and the direction of time becomes evident again.
\label{movie:S1}

\textbf{Entropy of state transitions in macroscopic active
matter}
\textbf{A} Experimental setup and reconstruction. 
This video shows the motorized spheres labelled with red markers confined in a circular boundary. Tracking both motions independently and reconstruction it subsequently allows the decomposition into translational and rotational components.
\textbf{B} Entropy and kinetic temperature during state transitions. 
he video highlights the evolution of entropy and kinetic temperature as the system transitions between different states. Discrete jumps in entropy correspond to these transitions, reflecting the emergence or reduction of degrees of freedom.
\textbf{C} Number variations reveal different states.
As the number of spheres in the system is varied, distinct phases emerge, demonstrating how reducing the number of spheres leads to different phase behaviors
\label{movie:S2}




\end{document}